\pdfoutput=1

\documentclass[11pt,twoside,a4paper,cmspaper,final,collab]{cms-tdr}
\def\svnVersion{493750P}\def\cmsCernNoTag{CERN-EP-2018-208}\def\cmsCernDate{\today}\def\cmsMessage{Published in Physics Letters B as \href{http://dx.doi.org/10.1016/j.physletb.2019.01.072}{\doi{10.1016/j.physletb.2019.01.072.}}}

\begin{document}\cmsNoteHeader{EXO-18-008}

\hyphenation{had-ron-i-za-tion}
\hyphenation{cal-or-i-me-ter}
\hyphenation{de-vices}
\RCS$HeadURL: svn+ssh://svn.cern.ch/reps/tdr2/papers/EXO-18-008/trunk/EXO-18-008.tex $
\RCS$Id: EXO-18-008.tex 487016 2019-01-23 15:43:49Z muahmad $

\newcommand{\qqZZ}{\ensuremath{\qqbar\to4\mu}\xspace}
\newcommand{\ZZ}{\ensuremath{\cPZ\gamma^{*}/\cPZ^{*}}\xspace}
\newcommand{\GsZs}{\ensuremath{\gamma^{*}/\cPZ^{*}}\xspace}
\newcommand{\ggZZ}{\ensuremath{\Pg\Pg\to4\mu}\xspace}
\newcommand{\ggH}{\ensuremath{\Pg\Pg\to\PH}\xspace}
\newcommand{\Zfmu}{\ensuremath{\cPZ\to4\PGm}\xspace}
\newcommand{\mfourmu}{\ensuremath{m_{4\PGm}}\xspace}
\newcommand{\Zprime}{\cPZ^{\prime}\xspace}
\newcommand{\mlplm}{\ensuremath{m_{\PGm^{+}\PGm^{-}}}\xspace}
\newcommand{\Zmm}{\mathrm{\cPZ\to\PGm\PGm}\xspace}
\newcommand{\usedLumi}{77.3\fbinv\xspace}
\newcommand{\minus}{\scalebox{0.75}[1.0]{$-$}}
\providecommand{\cmsTable}[1]{\resizebox{\textwidth}{!}{#1}}
\newlength\cmsTabSkip\setlength{\cmsTabSkip}{1ex}
\cmsNoteHeader{EXO-18-008}
\title{Search for an $L_{\mu}-L_{\Pgt}$ gauge boson using $\cPZ\to4\Pgm$ events in proton-proton collisions at $\sqrt{s} = 13$\TeV}

\date{\today}
\abstract{
A search for a narrow $\cPZpr$ gauge boson with a mass between 5 and $70\GeV$ resulting from an $L_{\Pgm}-L_{\Pgt}$ $U(1)$ local gauge symmetry is reported. Theories that predict such a particle have been proposed as an explanation of various experimental discrepancies, including the lack of a dark matter signal in direct-detection experiments, tension in the measurement of the anomalous magnetic moment of the muon, and reports of possible lepton flavor universality violation in \PB{} meson decays. A data sample of proton-proton collisions at a center-of-mass energy of 13\TeV is used, corresponding to an integrated luminosity of \usedLumi recorded in 2016 and 2017 by the CMS detector at the LHC. Events containing four muons with an invariant mass near the standard model $\cPZ$ boson mass are analyzed, and the selection is further optimized to be sensitive to the events that may contain $\cPZ \to \cPZpr\Pgm\Pgm\to4\Pgm$~decays. The event yields are consistent with the standard model predictions. Upper limits of $10^{-8}$--$10^{-7}$ at $95\%$ confidence level are set on the product of branching fractions $\mathcal{B}(\cPZ \to \cPZpr \Pgm \Pgm) \mathcal{B}(\cPZpr \to\Pgm \Pgm)$, depending on the $\cPZpr$ mass, which excludes a $\cPZpr$ boson coupling strength to muons above 0.004--0.3. These are the first dedicated limits on $L_{\Pgm}-L_{\Pgt}$ models at the LHC and result in a significant increase in the excluded model parameter space. The results of this search may also be used to constrain the coupling strength of any light $\cPZpr$  gauge boson to muons.}
\hypersetup{%
pdfauthor={CMS Collaboration},%
pdftitle={Search for an L[mu]-L[tau] gauge boson using Z to 4 mu events in proton-proton collisions at sqrt(s)=13 TeV},%
pdfsubject={CMS},%
pdfkeywords={CMS, physics, Zprime, muons}}

\maketitle

\section{Introduction}
The standard model (SM) of particle physics~\cite{SheldonL1961579,PhysRevLett.19.1264,SalamNobel} can not explain all experimental
observations to date and is, therefore, generally believed to be an incomplete theory.
Enlarging the SM gauge group to include an additional $U(1)$ symmetry is a simple and well-motivated extension~\cite{zprime1,zprime2,zprime3}, which leads to a prediction of a new
vector particle, a $\cPZpr$ boson.
In order for the extended gauge symmetry to be anomaly free, only certain generation-dependent couplings are allowed. The anomaly-free model we consider in this paper is the
$L_{\Pgm}-L_{\Pgt}$ gauge symmetry~\cite{PhysRevD.44.2118}, where $L_{\Pgm}$ and $L_{\Pgt}$ are the $\Pgm$ and $\PGt$ lepton numbers, respectively. The interaction
between the $\cPZpr$ and the second- and third-generation leptons can be described with the following Lagrangian~\cite{delAguila2015}:
\ifthenelse{\boolean{cms@external}}{
\begin{equation}\begin{split}
\label{eqn:lagrangian}
\mathcal{L}_{\cPZpr} = -g\cPZpr_{\Pgm}\Bigl(\overline{L}_{2}\gamma^{\Pgm}L_{2}+\overline{l}_{2}\gamma^{\Pgm}l_{2}\\
-\overline{L}_{3}\gamma^{\Pgm}L_{3}-\overline{l}_{3}\gamma^{\Pgm}l_{3}\Bigr),
\end{split}
\end{equation}
}
{
\begin{equation}
\label{eqn:lagrangian}
\mathcal{L}_{\cPZpr} = -g\cPZpr_{\Pgm}\left(\overline{L}_{2}\gamma^{\Pgm}L_{2}+\overline{l}_{2}\gamma^{\Pgm}l_{2}-\overline{L}_{3}\gamma^{\Pgm}L_{3}-\overline{l}_{3}\gamma^{\Pgm}l_{3}\right),
\end{equation}
}
where $g$ is an arbitrary dimensionless coupling to the SM left-handed and right-handed $\Pgm$ and $\PGt$ multiplets. These multiplets are:
\begin{equation}
\label{eqn:mutliplets}
L_{2} = \begin{pmatrix} \nu_{\Pgm} \\ \Pgm \end{pmatrix}_{L} \, \, , \, \, l_{2} = \Pgm_{R} \, \, , \, \, L_{3} = \begin{pmatrix} \nu_{\Pgt} \\ \Pgt \end{pmatrix}_{L} \, \, \text{and} \, \, l_{3} = \Pgt_{R}.
\end{equation}

Additional $U(1)$ gauge symmetries based on the difference in lepton family numbers are all anomaly free and require no new fermionic
particle content. The model based on gauging $L_{\Pgm}-L_{\Pgt}$ in particular is the least constrained experimentally,
since it is coupled only to second- and third-generation leptons. This model has gained popularity in
recent years~\cite{PhysRevD.64.055006,PhysRevD.89.095033,Harigaya2014,PhysRevD.90.035027,Altmannshofer2016,PhysRevD.93.015022,PhysRevD.96.015021} as an explanation for several
anomalous experimental measurements in particle physics.
These anomalies include the measurement of the anomalous muon magnetic moment by the Muon g$\minus 2$ Collaboration~\cite{PhysRevD.73.072003}, which can be explained for
certain values of the $\cPZpr$ mass and coupling strength~($g$)~\cite{PhysRevD.64.055006,Harigaya2014}. In addition, if the $\cPZpr$ mediates an interaction between
dark matter and ordinary matter, the bounds on the dark matter coupling strength from direct-detection experiments are less stringent~\cite{Altmannshofer2016,PhysRevD.90.035027} since the particular $\cPZpr$ considered here does not couple directly to quarks.
Finally, if additional interactions beyond the
minimal $L_{\Pgm}-L_{\Pgt}$ model are assumed, abnormalities in kinematic angular distributions and lepton flavor universality tests observed in
$\PB\to \PK^{*}\Pgm^{+}\Pgm^{-}$ decays~\cite{PhysRevLett.111.191801,PhysRevLett.113.151601} can be explained by this model, given its flavor
non-universal couplings~\cite{PhysRevD.89.095033,Altmannshofer2016}.

\begin{figure*}[!htb]
\centering
{ \includegraphics[width=0.45\textwidth]{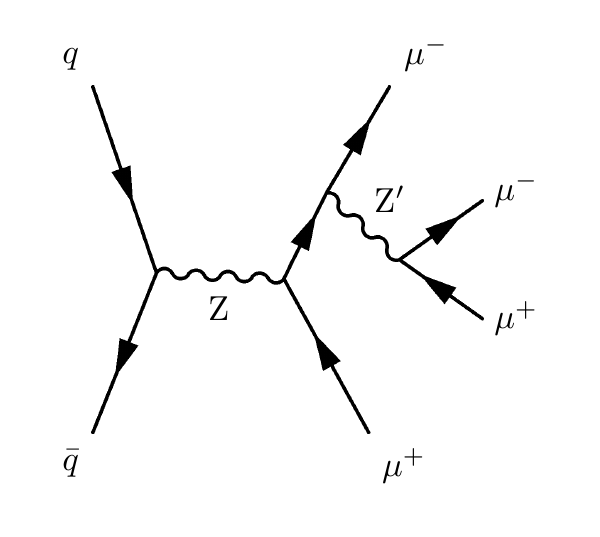}}
{ \includegraphics[width=0.45\textwidth]{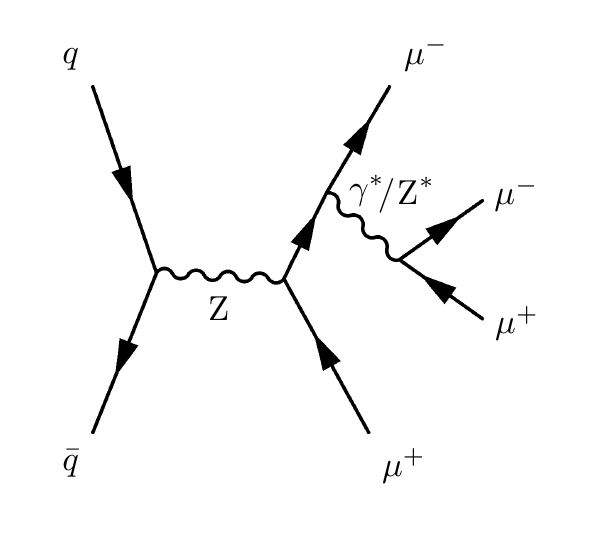}}
\caption{Leading order Feynman diagrams for the signal process (left) and the dominant background process (right), where in each diagram the four-muon final state
originates from annihilation.
\label{fig:feynman1}}
\end{figure*}

\begin{figure*}[!htb]
\centering
{ \includegraphics[width=0.45\textwidth]{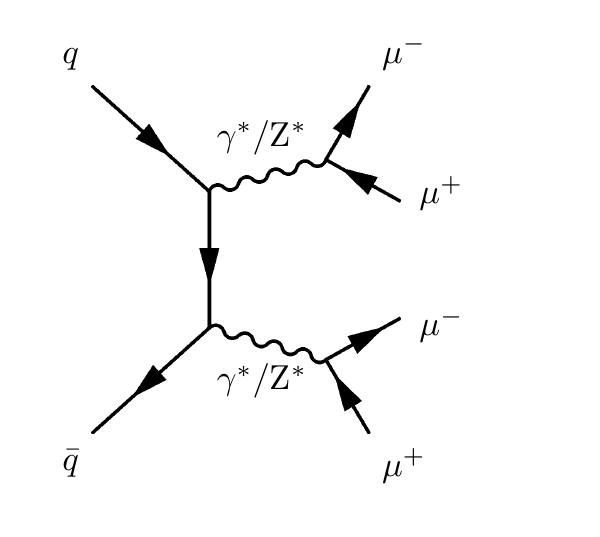}}
{ \includegraphics[width=0.45\textwidth]{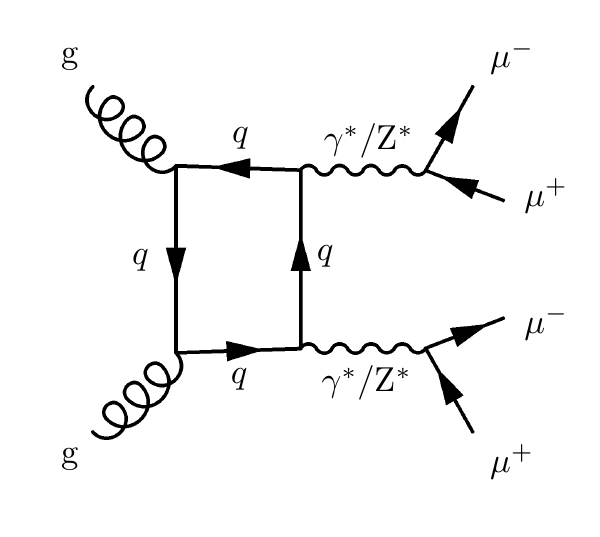}}
\caption{Leading order Feynman diagrams for the subdominant quark-initiated (left) and gluon-initiated (right) background processes, where in each diagram the four-muon final 
state originates from conversion.
\label{fig:feynman2}}
\end{figure*}

The $\cPZpr$ gauge boson associated with the putative $L_{\Pgm}-L_{\Pgt}$ gauge symmetry can be sought at the CERN LHC. Since the $\cPZpr$ couples only to second- and third-generation leptons ($\Pgm$, $\Pgngm$, $\Pgt$ and $\Pgngt$), it must be produced as a final state radiation product of a lepton originating from some other
physics process. The $\cPZ\to4\Pgm$ decay provides an extremely clean source of muons with excellent mass resolution. The resonant signal decay
$\cPZpr\to\Pgm\Pgm$ that may be present in $\cPZ\to4\Pgm$ decays further reduces the background contamination. There are two types of irreducible background  where the additional
dimuon originates from annihilation or conversion topologies, as described in  Ref.~\cite{Berends:1994pv}.
The Feynman diagrams in Fig.~\ref{fig:feynman1} (left) for the signal and in Fig.~\ref{fig:feynman1} (right) for the background are examples of the
 annihilation topology, while the diagrams in Fig.~\ref{fig:feynman2} are examples of the conversion topology.
The dominant background to the search comes from resonant $\cPZ$ production and decay to $4\Pgm$ from the annihilation diagram in Fig.~\ref{fig:feynman1} (right), while the
continuum background originating from the conversion diagrams in Fig.~\ref{fig:feynman2} are subdominant.
In the discussion of the analysis that follows, the signal and background processes originating
from the diagrams of Fig.~\ref{fig:feynman1} will hereafter be collectively referred to as the $\cPZ\to4\Pgm$ process since they have very similar kinematic properties. The
background processes originating from the diagrams in Fig.~\ref{fig:feynman1} (right) and Fig.~\ref{fig:feynman2} (left) will be collectively referred to as $\qqZZ$, and the
process originating from the diagram in Fig.~\ref{fig:feynman2} (right) will be referred to as $\ggZZ$.

The $\cPZ\to4\Pgm$  process has been studied by the ATLAS and CMS Collaborations~\cite{Chatrchyan2012,PhysRevLett.112.231806,Sirunyan2018} and constraints on
the $L_{\Pgm}-L_{\Pgt}$ parameter space have been derived. However, these measurements are not optimized for the presence of a $\cPZpr$ particle. In particular, they do
not utilize the fact that two of the four muons would form a resonant peak at the $\cPZpr$ mass, providing a means to reduce the dominant background by several orders of magnitude.
The subject of this paper is a dedicated counting experiment search with a final selection based on the reconstructed $\cPZpr$ candidate mass.


\section{The CMS detector}
\label{sec:detector}

A detailed description of the CMS detector, together with a definition of the coordinate system used and the relevant kinematic variables, can be found in Ref.~\cite{Chatrchyan:2008zzk}.
The central feature of the CMS apparatus is a superconducting solenoid of 6\unit{m} internal diameter, providing a magnetic field of 3.8\unit{T}.
Within the solenoid volume are a silicon pixel and strip tracker, a lead tungstate crystal electromagnetic calorimeter, and a brass and scintillator hadron calorimeter, each composed of a barrel and two endcap sections.
Forward calorimeters extend the pseudorapidity ($\eta$) coverage provided by the barrel and endcap detectors.
Muons are detected in gas-ionization chambers embedded in the steel flux-return yoke outside the solenoid.
The silicon tracker measures charged particles with $\abs{\eta}< 2.5$.
Muons are measured in the region $\abs{\eta}< 2.4$, with detection planes made using three technologies: drift tubes, cathode strip chambers, and resistive plate chambers.
Matching muons to tracks measured in the silicon tracker results in a relative transverse momentum ($\pt$) resolution for muons with $20\GeV <\pt < 100\GeV$ of 1.3--2.0\% in the
barrel ($\abs{\eta}< 0.9$) and better than 6\% in the endcaps ($\abs{\eta}>0.9$).
For charged hadrons, primarily used for the computation of muon isolation sums in this search, the track resolutions are typically 1.5\% in $\pt$ and 25--90 (45--150)$\mum$
in the transverse (longitudinal) impact parameter for transverse momentum between 1 and $10\GeV$ and $\abs{\eta} < 1.4$~\cite{TRK-11-001}.
The first level of the CMS trigger system~\cite{CMS-TRG-12-001}, composed of custom hardware processors, uses information from the calorimeters and muon detectors to select the
most interesting events in a fixed time interval of less than 4$\mus$. The high-level trigger (HLT) processor farm further decreases the event rate from around
100$\unit{kHz}$ to less than 1$\unit{kHz}$ before data storage.

\section{Data and simulated samples }
\label{sec:datasets}

This analysis makes use of proton-proton ($\Pp\Pp$) collision data recorded by the CMS detector in 2016 and 2017,
corresponding to an integrated luminosity of $\usedLumi$.
Collision events are selected by HLT algorithms that require the presence of one, two, or three muons passing
loose identification and isolation requirements. The main triggers used for this analysis
select a pair of muons where the minimal requirement for the transverse momentum with respect to the beam axis of the leading muon is $17\GeV$, while that for the subleading muon
is $8\GeV$. To maximize the signal acceptance, triggers requiring three muons with lower $\pt$ thresholds (12, 10 and $5\GeV$) and no isolation requirement are also used, as
are isolated single-muon triggers with the thresholds of $22\GeV$ and $27\GeV$ for 2016 and 2017 data taking, respectively. The overall trigger efficiency for
simulated signal events that pass
the full selection chain of this analysis (described in Section~\ref{sec:objects}) is greater than 99\%. The trigger efficiency is measured in data with a method based on
 the ``tag-and-probe'' technique~\cite{CMS:2011aa} using a sample
of $4\Pgm$ events collected by the single-muon triggers. Events with four muons have a negligible contamination from misidentified muons and therefore background
subtraction is not necessary.
Muons matched to the single-muon triggers are used as tags and the other three muons are
used as probes. The probe muons are then matched to the triggering muon objects from any of the one, two, or three muon triggers,
and the combined efficiency is extracted. The efficiency in data is found to be in agreement with the expectation from simulation.

Monte Carlo simulation samples for the $\cPZpr$ signal and for the background coming from the
$\qqZZ$ and $\ggZZ$ processes are used to optimize the event selection, evaluate the signal acceptance, and estimate the background rate and systematic uncertainties.
The signal is generated at leading order (LO) in perturbative quantum chromodynamics (pQCD) with \MGvATNLO (v2\_4\_2)~\cite{Alwall2014} together with the
Universal \textsc{FeynRules} Output (UFO) model from Ref.~\cite{delAguila2015}. The signal samples are generated with $m(\cPZpr)$ ranging from 5 to $70\GeV$
in steps of $5\GeV$. For $m(\cPZpr)$ below
$5\GeV$ nonprompt muons become a challenging background, and for $m(\cPZpr)$ above $70\GeV$ the $\cPZ$ boson starts to be produced off mass-shell, requiring a dedicated event selection. The
$\qqZZ$ process is generated at next-to-LO (NLO) in pQCD with \POWHEG 2.0~\cite{Nason:2004rx,Frixione:2007vw,Alioli:2010xd}, while
the $\ggZZ$ process is generated at LO with \MCFM~7.0~\cite{MCFM}.
The default set of parton distribution functions (PDFs) used in all simulations is NNPDF30\_nlo\_as\_0118~\cite{Ball2012153}.
The fully differential cross section for the $\qqZZ$~ process has been computed at next-to-NLO (NNLO)~\cite{Grazzini2015407},
and the appropriate NNLO/NLO correction factor $K$ of 1.03 at $m(4\Pgm)=m(\cPZ)$ is used to correct the \POWHEG sample.
The $\qqZZ$ background production is dominated by the conversion topology at $m(4\Pgm)\approx m(\cPZ)$, and therefore an analogous NNLO/LO $K$ factor of 1.29 is used to
correct the signal process that originates from the same topology. The $\ggZZ$ process contributes at NNLO in pQCD and is corrected by a $K$ factor of
2.4~\cite{Bonvini:1304.3053,Melnikov:2015laa,Li:2015jva,Passarino:1312.2397v1,Catani:2007vq,Grazzini:2008tf,Grazzini:2013mca}.

After the final selection, described in Section~\ref{sec:eventsel}, the $\ggZZ$ background contribution is typically less than 1\% (and at most 7\%) of
the $\qqZZ$ contribution. These simulations have been found to provide an accurate description of $4\Pgm$ events in data by several previous studies~\cite{CMSH4lLegacy,CMSH4lFiducial8TeV,HIG16041,Sirunyan2018}. All the generated events are interfaced with \PYTHIA 8.212~\cite{Sjostrand2015159} tune CUETP8M1~\cite{Khachatryan:2015pea} to simulate multiple
parton interactions, the underlying event, and the fragmentation and hadronization effects. The generated events are processed through a detailed simulation of the CMS detector based on \GEANTfour~\cite{Agostinelli:2002hh,GEANT} and reconstructed with the same algorithms that are used for the data.
The simulated events include overlapping $\Pp\Pp$ interactions (pileup) and have been reweighted so that the distribution of the number of interactions per LHC bunch crossing in simulation matches that  observed in data.

\section{Object reconstruction}
\label{sec:objects}

The techniques of the object reconstruction and event selection are based largely on Refs.~\cite{CMSH4lLegacy,CMSH4lFiducial8TeV,HIG16041,Sirunyan2018}.
Event reconstruction is based on the particle-flow (PF) algorithm~\cite{Sirunyan:2017ulk}, which exploits information from all the CMS subdetectors to identify
and reconstruct individual particles in the event. Higher-level observables, such as muon isolation quantities, are built from
the PF candidates classified as charged hadrons, neutral hadrons, photons, electrons, or muons.

Muons are reconstructed within the geometrical acceptance $\abs{\eta} < 2.4$ by combining information from the silicon tracker and the muon
system~\cite{Sirunyan:2018fpa}, and are required to satisfy $\PT > 5\GeV$.
The inner and outer tracks are matched using either an outside-in algorithm, starting from a track in the muon system,
or an inside-out algorithm, starting from a track in the silicon tracker.
In the latter case, some very low-$\pt$ muons that may not have sufficient energy to penetrate the entire muon system are also collected by
considering tracks that match track segments in only one or two planes of the muon system.
Muons are identified from the reconstructed muon track candidates by applying minimal requirements on the inner and outer tracks, taking into account
their compatibility with small energy deposits in the calorimeters~\cite{Sirunyan:2017ulk}.

Muons originating from nonprompt decays of hadrons are suppressed by
requiring each muon track to have the ratio between its impact parameter in three dimensions, computed with respect to the chosen primary vertex position,
and its uncertainty to be less than 4.
The primary $\Pp\Pp$ interaction vertex is taken to be the reconstructed vertex with the largest value of summed $\pt^2$ of jets and associated missing transverse
momentum, calculated from the tracks assigned to the vertex, where the jet finding algorithm is taken from Refs.~\cite{Cacciari:2008gp,Cacciari:2011ma} and the missing
transverse momentum is taken as the negative vector sum of the $\pt$ of the jets.

A relative isolation requirement of $\mathcal{I}^{\PGm}<0.35$ is imposed
to discriminate between prompt muons from $\cPZ$ boson decays and those arising from electroweak decays of hadrons within jets,
where the relative isolation is defined as

\ifthenelse{\boolean{cms@external}}{
\begin{equation}\begin{split}
\label{eqn:pfiso}
\mathcal{I}^{\PGm} \equiv& \Biggl( \sum \PT^\text{charged} +
                                 \max\Bigl[ 0, \sum \PT^\text{neutral}
                                 +\\
                                  &\sum \PT^{\Pgg}
                                 - \PT^\mathrm{PU} \Big] \Biggr)
                                 / \PT^{\Pgm}.
\end{split}
\end{equation}
}{
\begin{equation}
\label{eqn:pfiso}
\mathcal{I}^{\Pgm} \equiv \Big( \sum \PT^\text{charged} +
                                 \max\left[ 0, \sum \PT^\text{neutral}
                                 +
                                  \sum \PT^{\Pgg}
                                 - \PT^\mathrm{PU} \right] \Big)
                                 / \PT^{\Pgm}.
\end{equation}
}

The isolation sums involved are all restricted to PF candidates within a volume bounded by a
cone of angular radius  $\Delta R=0.3$ around the muon direction at the
primary vertex, where the angular distance between two particles
$i$ and $j$ is $\Delta R(i,j) = \sqrt{\smash[b]{(\eta^i-\eta^j)^{2} + (\phi^i-\phi^j)^{2}}}$ and $\phi$ is the azimuthal angle in radians.
The quantity $\sum \PT^\text{charged}$ is the scalar sum of the transverse momenta of charged hadrons originating from
the chosen primary vertex of the event. Charged hadrons are associated with charged particle tracks assigned neither to electrons nor to muons.
The quantities $\sum \PT^\text{neutral}$ and $\sum \PT^{\Pgg}$ are the
scalar sums of the transverse momenta for neutral hadrons and photons, respectively.
Energy deposits from pileup interactions, $\PT^\text{PU}$, are subtracted to make the isolation variable less sensitive to the number of pileup interactions.
Here, we define $\PT^\mathrm{PU} \equiv 0.5 \sum_i \PT^{\mathrm{PU}, i}$, where $i$ runs over the momenta of the charged hadron PF candidates not originating from the primary vertex and the factor of 0.5 accounts for the different fractions of charged and neutral particles in the cone.

An algorithm is used to recover the final-state radiation (FSR) photons from muons.
Photons reconstructed by the PF algorithm within $\abs{\eta_\Pgg} < 2.4$ are required to satisfy
$\PT^{\Pgg} > 2\GeV$ and $\mathcal{I}^{\Pgg}< 1.8$. The photon relative isolation $\mathcal{I}^{\Pgg}$ is defined as
for the muons in Eq.~(\ref{eqn:pfiso}). Every FSR candidate is associated with the closest selected muon in the event,
and we require FSR candidates to satisfy $\Delta R(\cPgg,\Pgm)/(\PT^{\Pgg})^{2}<0.012\GeV^{-2}$ and $\Delta R(\cPgg,\Pgm)<0.5$.
Finally, for every muon we retain the FSR candidate, if any, with the lowest $\Delta R(\cPgg,\Pgm)/(\PT^{\Pgg})^{2}$. About 5\% of signal events are found to have one FSR photon
attached. Any selected FSR photons are excluded from the corresponding muon isolation computation.

The decay products of known dimuon resonances ($\PJGy$ meson, $\cPZ$ boson) are used to calibrate the muon momentum scale and resolution in bins of $\pt$ and $\eta$.
Muon momenta are calibrated using a Kalman filter approach~\cite{Fruhwirth:1987fm}.
A tag-and-probe technique~\cite{CMS:2011aa,Chatrchyan:2012xi}
is used to measure the efficiency of the reconstruction and selection
for prompt muons in several bins of $\pt$ and $\eta$.
The difference between the efficiencies measured in simulation and data, which on average is 1.2\% per muon,
is used to correct the selection efficiency in the simulated samples. The combined muon reconstruction and identification efficiency for signal events, including these corrections, is about 92\% per muon.

\section{Event selection}
\label{sec:eventsel}
Events are required to contain at least four well-identified and isolated muons, with at least two muons required to have $\pt > 10\GeV$ and at least one to have $\pt > 20\GeV$.
The four selected muons must have zero net charge.
Dimuon candidates are formed from muon pairs of opposite sign ($\Pgm^{+}\Pgm^{-}$) and are required to pass $4\GeV < \mlplm  < 120\GeV$.
All recovered FSR photon candidates are included in the invariant mass computation.
The dimuon candidates are then combined into $\cPZ\to4\Pgm$ candidates. We define $\cPZpr_1$ to be the dimuon candidate with the highest invariant mass,
and $\cPZpr_2$ as the other one.
In events with more than four muons where several $\cPZ\to4\Pgm$ candidates have the same $m(\cPZpr_1)$, the $\cPZpr_2$ candidate formed from the two muons with the highest
scalar sum of $\pt$ is chosen.

To be considered for the analysis, $\cPZ\to4\Pgm$ candidates have to pass a set of kinematic requirements.
The $\cPZpr_1$ invariant mass must be larger than $12\GeV$ and all muons must be separated in angular space by at least $\Delta R(\Pgm_i, \Pgm_j) > 0.02$.
To further suppress events with muons originating from hadron decays in jet fragmentation or from the decay of low-mass hadronic resonances, all
four opposite sign muon pairs that can be constructed with the four muons are required to satisfy $m_{\Pgm^{+}\Pgm^{-}} > 4\GeV$, where selected FSR photons are
disregarded in the invariant mass computation.
Finally, the four-muon invariant mass $m(4\Pgm)$ must be between 80 and 100\GeV.
The $\cPZpr$ candidate is most often reconstructed as $\cPZpr_2$ for $m(\cPZpr)<42.65\GeV$ and as $\cPZpr_1$ for $m(\cPZpr)>42.65\GeV$. The search is a counting
experiment with a sliding mass window, and a final selection made on either $m(\cPZpr_1)$ or $m(\cPZpr_2)$ values, depending on the $\cPZpr$ mass hypothesis. The exclusion limit
for $m(\cPZpr)=42.65\GeV$ using either $m(\cPZpr_2)$ or $m(\cPZpr_1)$ as an observable is about the same.
For $m(\cPZpr)<42.65\GeV$, $m(\cPZpr_2)$ is required to be within 2\% of the $m(\cPZpr)$. While for $m(\cPZpr)>42.65$~$\GeV$, the same requirement is applied on $m(\cPZpr_1)$.
The search window size of 2\% was chosen to simultaneously optimize the expected significance and exclusion limit for different $\cPZpr$ mass hypotheses.
The efficiency of this requirement is directly related to the efficiency of selecting the correct $\cPZpr$ candidate and varies with $\cPZpr$ mass. It is found to
be about 63\% for $m(\cPZpr)=5\GeV$, 25\% for $m(\cPZpr)=40\GeV$, and 67\% for $m(\cPZpr)=70\GeV$.
The low efficiency for $m(\cPZpr)\approx m_{\cPZ}/2$ is due the combinatoric ambiguity in selecting the correct $\cPZpr$ candidate from the four possible dimuon
candidates.
The selection is, however, still extremely beneficial, as it eliminates approximately 99.8\% of the SM $\GsZs$ background for $m(\cPZpr)=40\GeV$.
Additional backgrounds to the signal that can arise
from processes in which heavy-flavor jets produce secondary muons, and
from processes in which decays of heavy-flavor hadrons or nonprompt decays of
light mesons within jets are misidentified as prompt muons, are found to be negligible after the final event selection.

\section{Systematic uncertainties}
\label{sec:systematics}
Experimental uncertainties that equally affect the signal and background estimations include the uncertainty in the integrated luminosity measurement of
$2.5$\%~\cite{CMS-PAS-LUM-17-001} and $2.3$\%~\cite{CMS-PAS-LUM-17-004} for the 2016 and 2017 data sets, respectively, and the uncertainty in the
muon reconstruction, identification, and isolation efficiency ($4.9$\% on the overall event yield).
An uncertainty in the signal and background yields due to the muon momentum scale is determined using $\cPZ\to4\Pgm$ events in data and simulation and found to be
negligible ($0.1$\%).
An uncertainty in the signal and background yields of $2.0$\% coming from the
determination of the muon momentum resolution is obtained by smearing the dimuon mass resolution by $20$\%~\cite{HIG16041} with respect to the nominal resolution and recomputing the expected yields.
The uncertainties due to the finite sizes of the simulated samples amount to $3.0$\% for the background estimation and $1.4$\% for the signal estimation.

Theoretical uncertainties that affect both the signal and background estimations
include uncertainties in the finite-order perturbative calculations and the choice of the PDF set.
The uncertainties arising from finite-order perturbative calculations are estimated by varying the renormalization and factorization scales between
$0.5$ and $2.0$ times their nominal value, while keeping their ratios between $0.5$ and $2.0$.
This uncertainty is found to be $3.5$ ($3.9$)\% for the $\qqZZ$ ($\ggZZ$) process and is taken to be correlated between the signal and the dominant $\qqZZ$ background process.
The uncertainty due to missing electroweak corrections in the
region $m(4\Pgm)\approx m(\cPZ)$ is expected to be small compared to the uncertainties in the pQCD calculation.
Following Ref.~\cite{Bonvini:1304.3053} and taking into account differences in selection, an additional uncertainty of $10$\% in the $K$ factor used for
the $\ggZZ$ prediction described in Section~\ref{sec:datasets} is applied to account for the fact that the $K$ factor was computed for the $\ggH$ process.
The uncertainty from the PDF set is determined following the PDF4LHC recommendations~\cite{Butterworth:2015oua} and is found to be $3.1$ ($3.5$)\% for the $\qqZZ$ ($\ggZZ$) process.
This uncertainty is also taken to be correlated between the signal and the dominant $\qqZZ$ background process.

To estimate the effect of the interference between the signal
and background processes, three types of samples are generated
using \MGvATNLO (v2\_4\_2): $\Pp\Pp\to4\Pgm$ (inclusive), $\Pp\Pp\to\cPZpr\Pgm\Pgm~\to4\Pgm$ (signal only), and $\Pp\Pp\to4\Pgm$ (background only). In all $g$ is varied
from $0.01$ to $0.50$, which corresponds to relative widths of less than 2\% in the model considered. The inclusive sample contains background, signal, and
interference contributions. The effect of the interference on the normalization of the signal is estimated by taking the difference of the inclusive sample cross section
and the sum of the cross sections of the signal and background samples. This difference is at most $5.0$\% after the final event selection, and an
additional $5.0$\% uncertainty in the signal yield is applied to account for this effect.

The combined systematic uncertainties in the background and signal yields are about 8\% and 10\%, respectively.

\section{Results}
\label{sec:results}
\begin{figure}[!htb]
\centering
\includegraphics[height=0.45\textwidth]{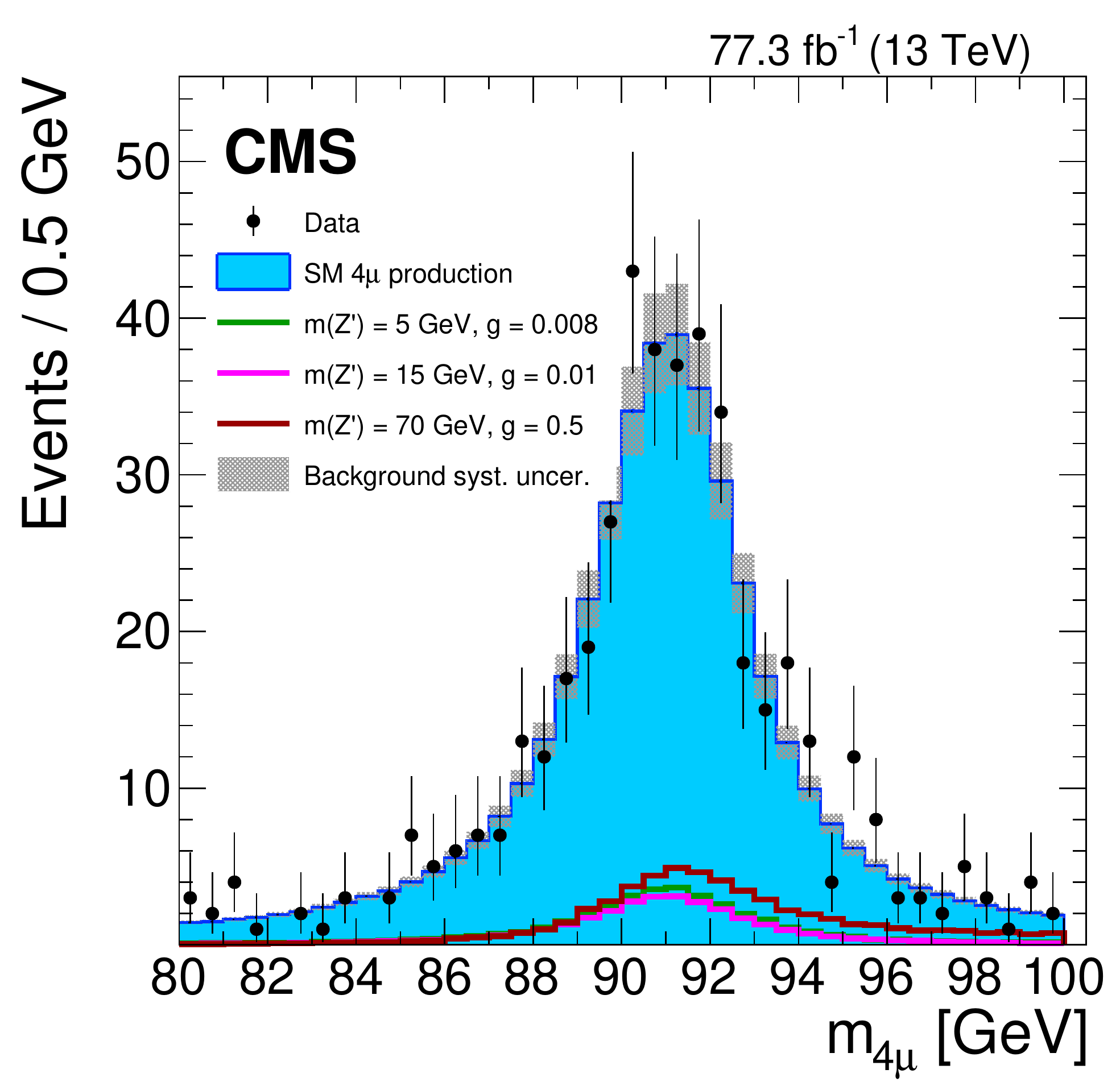}
\caption{Distribution of the reconstructed four-muon invariant mass $\mfourmu$ in the full mass range and a comparison to the predicted $\qqbar/\Pg\Pg\to4\Pgm$
background. The blue histogram represents the expected SM~$4\Pgm$~background distribution
and the gray band shows the systematic uncertainty in its prediction. For illustration, three $\cPZpr$ signal hypotheses with different masses and coupling
strengths are shown by colored lines.
\label{fig:mass4mu}}
\end{figure}

\begin{figure*}[!htbp]
\centering
{ \includegraphics[width=0.45\textwidth]{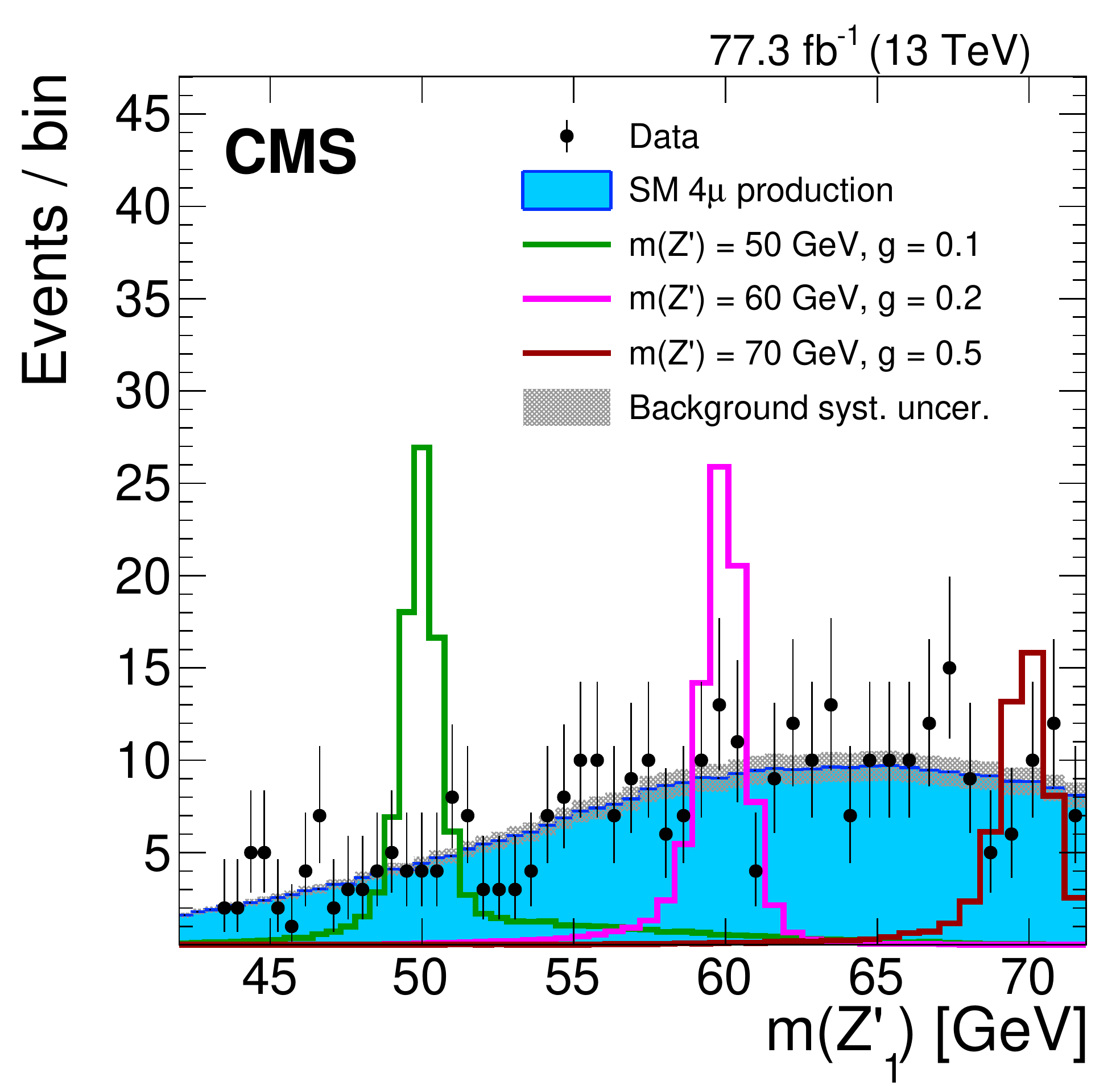}}
{ \includegraphics[width=0.45\textwidth]{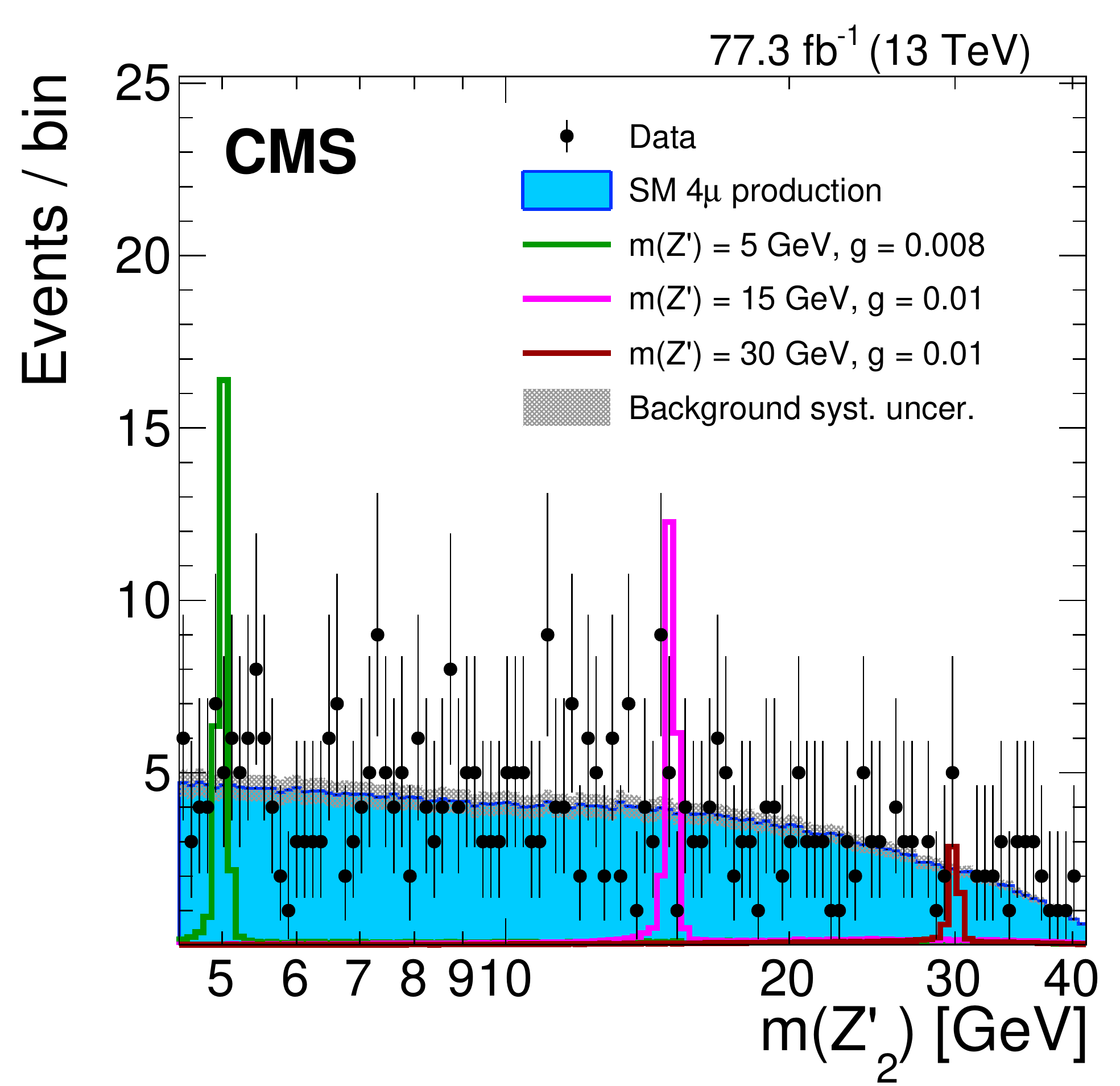}}
\caption{Distributions of the reconstructed $m(\cPZpr_1)$ and $m(\cPZpr_2)$ observables and a comparison to the predicted $\qqbar/\Pg\Pg\to4\Pgm$ background. The
variable bin width has been chosen according to the expected mass resolution. The blue histogram represents the expected SM~$4\Pgm$~background distributions and the gray
band shows the systematic uncertainty in its prediction. For illustration, three $\cPZpr$ signal hypotheses with different masses and coupling strengths
are also shown by colored lines.
\label{fig:massZ1Z2}}
\end{figure*}

The number of candidates observed in data and the expected yields for the backgrounds and the different $\cPZpr$ signals after the full event
selection are reported in Table~\ref{tab:EventYieldsFull}.
The reconstructed four-muon invariant mass distributions are shown in Fig.~\ref{fig:mass4mu} and compared with the expectations from signal
and background processes. Fig.~\ref{fig:massZ1Z2} shows the reconstructed $m(\cPZpr_1)$ and $m(\cPZpr_2)$ distributions.

\begin{table*}[!htb]
\centering
\topcaption{The numbers of expected background and signal events and the numbers of observed candidate events after the full selection with $80\GeV<\mfourmu<100\GeV$.
The signal and $\qqbar/\Pg\Pg\to4\Pgm$ background rates are both estimated from simulation. The signal predictions are reported with systematic uncertainties only, while
the background predictions are reported with statistical and systematic uncertainties, respectively. Also shown are the numbers of expected background and signal events and
the numbers of observed candidate events in the relevant mass windows for three $m(\cPZpr)$ hypotheses.
The values of the coupling strengths are chosen for the purpose of illustration.
\label{tab:EventYieldsFull}}
\cmsTable{
\begin{tabular}{cccccc}
 \hline

\multirow{2}{*}{} & \multirow{2}{*}{Background} & $m(\cPZpr)=5\GeV$ & $m(\cPZpr)=15\GeV$ & $m(\cPZpr)=70\GeV$ & Observed  \\
 & & $g=0.008$ & $g=0.01$ & $g=0.5$ & Data \\ \hline
$80\GeV<\mfourmu<100\GeV$ & $423.0\pm20.6\pm33.4$ & $37.1\pm3.7$ & $31.4\pm3.1$ & $53.8\pm5.4$ & 441 \\[\cmsTabSkip]
$4.9\GeV<m(\cPZpr_{2})<5.1\GeV$   & $9.2\pm3.0\pm0.7$ & $23.3\pm2.3$ & $\NA$ & $\NA$ & 13 \\
$14.7\GeV<m(\cPZpr_{2})<15.3\GeV$ & $7.7\pm2.8\pm0.6$ & $\NA$ & $18.9\pm1.9$ & $\NA$ & 6 \\
$68.6\GeV<m(\cPZpr_{1})<71.4\GeV$ & $34.9\pm5.9\pm2.8$ & $\NA$ & $\NA$ & $36.0\pm3.6$ & 35 \\
Predicted $\sigma\times\mathcal{B}\,\,[\text{fb}]$ & \NA & $9.6$ & $3.0$& $12$ & \NA\\
\hline
\end{tabular}
}
\end{table*}

\begin{figure}[!htb]
\centering
{ \includegraphics[width=0.49\textwidth]{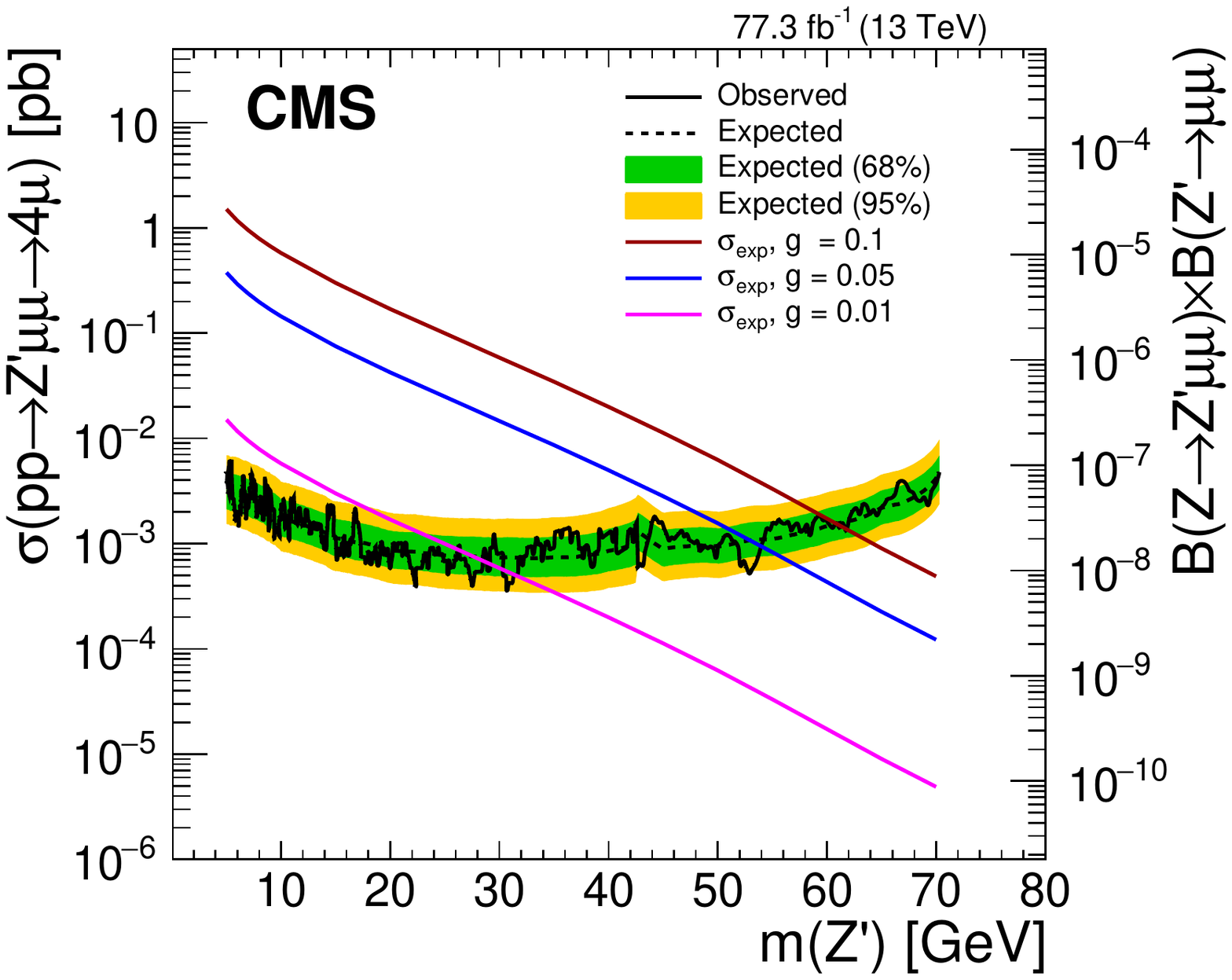}}
\caption{Expected and observed 95\% $\CL$ limits on the product of the $\cPZpr\mu\mu$~production cross section and branching fraction (left y-axis)
and $\mathcal{B}(\cPZ \to \cPZpr\Pgm\Pgm) \mathcal{B}(\cPZpr\to\Pgm\Pgm)$ (right y-axis). The dashed black curve is the expected upper limit, with one
and two standard-deviation bands shown in green and yellow, respectively. The solid black curve is the observed upper limit. The colored lines show the predicted cross section
times branching fraction (left y-axis) and $\mathcal{B}(\cPZ \to \cPZpr\Pgm\Pgm) \mathcal{B}(\cPZpr\to\Pgm\Pgm)$ (right y-axis) as a function
of $m(\cPZpr)$ for three different coupling strengths, chosen for illustration. The $\mathcal{B}(\Zprime\to\Pgm\Pgm)$ is taken to be $1/3$ to derive the theoretical predictions.
\label{fig:limits_XS}}
\end{figure}

In all cases, the observed distributions agree with the expectations within the assigned uncertainties. Upper limits at 95\% confidence level ($\CL$)
are derived on the product of the $\Zprime\Pgm\Pgm$~production cross section and the branching fraction $\mathcal{B}(\cPZpr\to\Pgm\Pgm)$ using the $\CLs$ method~\cite{CLS2,CLS1} with the
test statistic described in Ref.~\cite{LHC-HCG}, in the asymptotic approximation~\cite{Cowan:2010js}.
The asymptotic approximation was verified to be valid by computing limits with the full $\CLs$ method using
pseudo-experiments for several $m(\cPZpr)$ hypotheses.
A linear interpolation of the expected event yields between generated signal simulation samples is assumed in the limit calculations.
Systematic uncertainties are incorporated into the likelihood as nuisance parameters with log-normal probability distributions. Due to the low number of events passing the
final selection, the statistical uncertainty is always larger than 22\% within the entire $m(\cPZpr)$ search region, and dominates the sensitivity of this analysis. These limits are
shown in Fig.~\ref{fig:limits_XS}. The upper limits on
the $\mathcal{B}(\cPZ \to \cPZpr\Pgm\Pgm) \mathcal{B}(\cPZpr\to\Pgm\Pgm)$ are also shown. For the derivation of branching fraction
limits, the $\cPZ$~boson production cross section prediction computed at NNLO in pQCD with the program
FEWZ~2.1~\cite{PhysRevLett.96.231803,Gavin:2010az,Gavin:2012sy} is used.

\begin{figure}[!htb]
\centering
{ \includegraphics[width=0.45\textwidth]{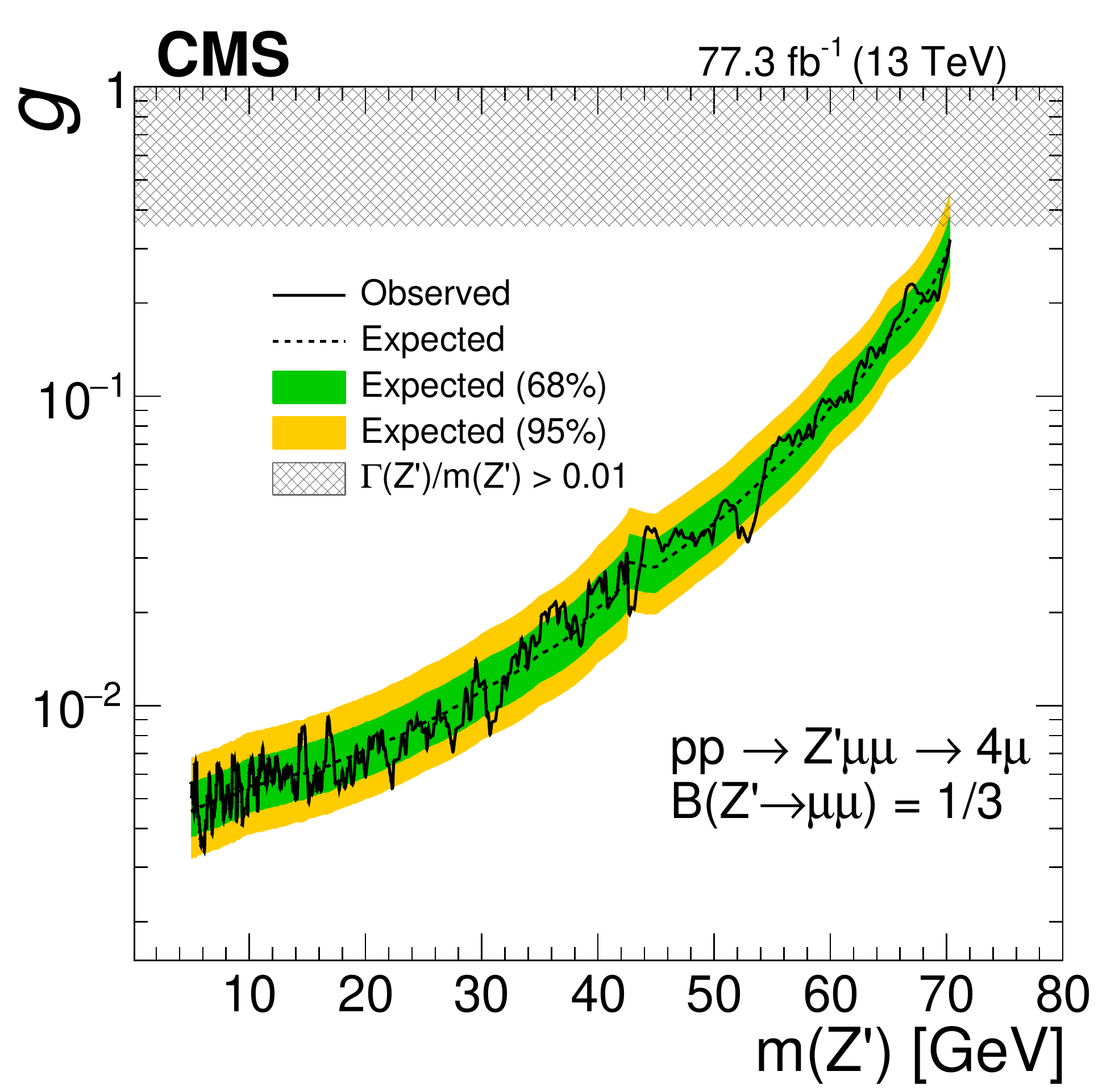}} \\
{ \includegraphics[width=0.45\textwidth]{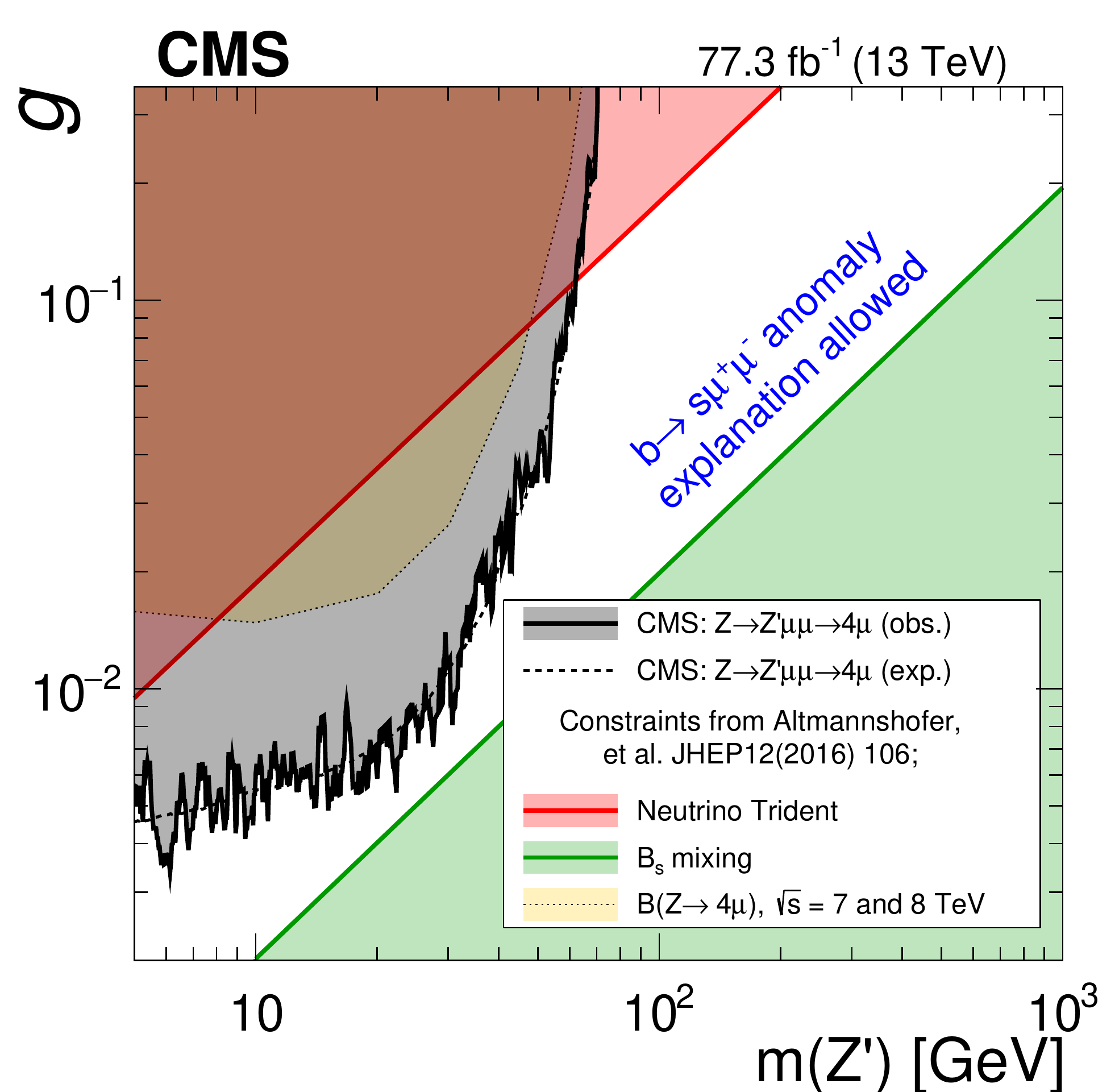}}
\caption{Top: Expected and observed 95\% $\CL$ limits on the gauge coupling strength $g$ as a function of $m(\cPZpr)$. The dashed black curve is the expected upper limit, with one
and two standard-deviation bands shown in green and yellow, respectively. The solid black curve is the observed upper limit. The $\mathcal{B}(\Zprime\to\Pgm\Pgm) = 1/3$ is
used to derive the upper limits. The hatched area shows the region where the narrow width approximation is no longer valid.
Bottom: comparison with other experiments sensitive to the same parameter space, with shaded regions being excluded as described in the text. These three constraints are
adapted from Ref.~\cite{Altmannshofer2016}.
\label{fig:limits}}
\end{figure}

Upper limits are also derived on the gauge coupling strength $g$ and compared to other experimental
constraints, shown in Fig.~\ref{fig:limits}.
These limits assume the $\mathcal{B}(\cPZpr\to\Pgm\Pgm)$ is equal to $1/3$ as in the minimal $L_{\Pgm}-L_{\Pgt}$ model with equal left- and right-handed
coupling strengths, and the additional constraints are adapted from
Ref.~\cite{Altmannshofer2016}. The mass of the dark matter candidate in the model from Ref.~\cite{Altmannshofer2016} is assumed to be much larger than the
largest $\cPZpr$ mass considered and the gauge coupling strengths to other particles, such as $\PQb$- and $\PQs$-quarks, are taken to be much smaller than the
coupling strength to leptons so that $\mathcal{B}(\cPZpr\to\Pgm\Pgm)$ is constant.
The natural width of the $\cPZpr$ is also assumed to be less than the
detector resolution,
which is a valid approximation in the minimal $L_{\Pgm}-L_{\Pgt}$ model when $g^2/4\pi<0.01$.
The shaded yellow region shows constraints derived in Ref.~\cite{Altmannshofer2016} from the ATLAS $\mathcal{B}(\cPZ\to4\Pgm)$ measurement at $\sqrt{s}=7$ and
$8\TeV$~\cite{PhysRevLett.112.231806}.
The shaded red region is excluded by the measurement of the so-called neutrino trident cross section by the CCFR Collaboration~\cite{PhysRevLett.66.3117,PhysRevLett.113.091801}.
The green region is excluded by a global analysis of $\PBs$ mixing measurements performed in Ref.~\cite{Altmannshofer2016}.
The region in between those two constraints and for $m(\cPZpr)>10\GeV$ is a candidate region to explain the LHCb B decay anomalies~\cite{PhysRevLett.111.191801,PhysRevLett.113.151601}.
It is important to note that in order to explain these anomalies, additional assumptions on the couplings of the $\cPZpr$ boson to $\PQb$- and $\PQs$-quarks are required,
and the constraints from $\PBs$ mixing measurements are therefore not generally applicable to the minimal $L_{\Pgm}-L_{\Pgt}$ model.
 It can be seen that this search is able to exclude a significant portion of the previously allowed parameter space.

\section{Summary}
\label{sec:summary}
A search for a $\cPZpr$ gauge boson resulting from an $L_{\Pgm}-L_{\Pgt}$ $U(1)$ local gauge symmetry is presented, based on data from proton-proton
collisions at $\sqrt{s}=13\TeV$ corresponding to an integrated luminosity of $\usedLumi$ recorded in 2016 and 2017 by the CMS detector at the LHC. Events with four muons
having an invariant mass near the mass of the standard model $\cPZ$ boson are selected, and the search sensitivity is optimized for the presence of
$\cPZ \to \cPZpr\Pgm\Pgm\to4\Pgm$~decays. The search places strong constraints on theories that attempt to explain various experimental anomalies including the lack of a dark
matter signal in direct-detection experiments, tension in
the measurement of the anomalous magnetic moment of the muon, and reports of possible lepton flavor universality violation in \PB{} meson decays.
The event yields are consistent with the standard model expectations. Upper limits of $10^{-8}$--$10^{-7}$ at 95\% confidence level are set on
the product of branching fractions $\mathcal{B}(\cPZ \to \cPZpr\Pgm\Pgm) \mathcal{B}(\cPZpr\to\Pgm\Pgm)$, depending on the $\cPZpr$ mass, which
excludes a $\cPZpr$ boson coupling strength to muons above 0.004--0.3. These are the first dedicated limits on $L_{\Pgm}-L_{\Pgt}$ models at the LHC and result in a significant
increase in the excluded model parameter space.

\begin{acknowledgments}
We congratulate our colleagues in the CERN accelerator departments for the excellent performance of the LHC and thank the technical and administrative staffs at CERN and at other CMS institutes for their contributions to the success of the CMS effort. In addition, we gratefully acknowledge the computing centers and personnel of the Worldwide LHC Computing Grid for delivering so effectively the computing infrastructure essential to our analyses. Finally, we acknowledge the enduring support for the construction and operation of the LHC and the CMS detector provided by the following funding agencies: BMWFW and FWF (Austria); FNRS and FWO (Belgium); CNPq, CAPES, FAPERJ, and FAPESP (Brazil); MES (Bulgaria); CERN; CAS, MoST, and NSFC (China); COLCIENCIAS (Colombia); MSES and CSF (Croatia); RPF (Cyprus); SENESCYT (Ecuador); MoER, ERC IUT, and ERDF (Estonia); Academy of Finland, MEC, and HIP (Finland); CEA and CNRS/IN2P3 (France); BMBF, DFG, and HGF (Germany); GSRT (Greece); NKFIA (Hungary); DAE and DST (India); IPM (Iran); SFI (Ireland); INFN (Italy); MSIP and NRF (Republic of Korea); LAS (Lithuania); MOE and UM (Malaysia); BUAP, CINVESTAV, CONACYT, LNS, SEP, and UASLP-FAI (Mexico); MBIE (New Zealand); PAEC (Pakistan); MSHE and NSC (Poland); FCT (Portugal); JINR (Dubna); MON, RosAtom, RAS and RFBR (Russia); MESTD (Serbia); SEIDI, CPAN, PCTI and FEDER (Spain); Swiss Funding Agencies (Switzerland); MST (Taipei); ThEPCenter, IPST, STAR, and NSTDA (Thailand); TUBITAK and TAEK (Turkey); NASU and SFFR (Ukraine); STFC (United Kingdom); DOE and NSF (USA).

\hyphenation{Rachada-pisek} Individuals have received support from the Marie-Curie program and the European Research Council and Horizon 2020 Grant, contract No. 675440 (European Union); the Leventis Foundation; the A. P. Sloan Foundation; the Alexander von Humboldt Foundation; the Belgian Federal Science Policy Office; the Fonds pour la Formation \`a la Recherche dans l'Industrie et dans l'Agriculture (FRIA-Belgium); the Agentschap voor Innovatie door Wetenschap en Technologie (IWT-Belgium); the F.R.S.-FNRS and FWO (Belgium) under the ``Excellence of Science - EOS'' - be.h project n. 30820817; the Beijing Municipal Science \& Technology Commission, No. Z181100004218003; the Ministry of Education, Youth and Sports (MEYS) of the Czech Republic; the Lend\"ulet (``Momentum'') Program and the J\'anos Bolyai Research Scholarship of the Hungarian Academy of Sciences, the New National Excellence Program \'UNKP, the NKFIA research grants 123842, 123959, 124845, 124850 and 125105 (Hungary); the Council of Science and Industrial Research, India; the HOMING PLUS program of the Foundation for Polish Science, cofinanced from European Union, Regional Development Fund, the Mobility Plus program of the Ministry of Science and Higher Education, the National Science Center (Poland), contracts Harmonia 2014/14/M/ST2/00428, Opus 2014/13/B/ST2/02543, 2014/15/B/ST2/03998, and 2015/19/B/ST2/02861, Sonata-bis 2012/07/E/ST2/01406; the National Priorities Research Program by Qatar National Research Fund; the Programa Estatal de Fomento de la Investigaci{\'o}n Cient{\'i}fica y T{\'e}cnica de Excelencia Mar\'{\i}a de Maeztu, grant MDM-2015-0509 and the Programa Severo Ochoa del Principado de Asturias; the Thalis and Aristeia programs cofinanced by EU-ESF and the Greek NSRF; the Rachadapisek Sompot Fund for Postdoctoral Fellowship, Chulalongkorn University and the Chulalongkorn Academic into Its 2nd Century Project Advancement Project (Thailand); the Welch Foundation, contract C-1845; and the Weston Havens Foundation (USA).
\end{acknowledgments}

\bibliography{auto_generated}

\cleardoublepage \appendix\section{The CMS Collaboration \label{app:collab}}\begin{sloppypar}\hyphenpenalty=5000\widowpenalty=500\clubpenalty=5000\vskip\cmsinstskip
\textbf{Yerevan Physics Institute, Yerevan, Armenia}\\*[0pt]
A.M.~Sirunyan, A.~Tumasyan
\vskip\cmsinstskip
\textbf{Institut f\"{u}r Hochenergiephysik, Wien, Austria}\\*[0pt]
W.~Adam, F.~Ambrogi, E.~Asilar, T.~Bergauer, J.~Brandstetter, M.~Dragicevic, J.~Er\"{o}, A.~Escalante~Del~Valle, M.~Flechl, R.~Fr\"{u}hwirth\cmsAuthorMark{1}, V.M.~Ghete, J.~Hrubec, M.~Jeitler\cmsAuthorMark{1}, N.~Krammer, I.~Kr\"{a}tschmer, D.~Liko, T.~Madlener, I.~Mikulec, N.~Rad, H.~Rohringer, J.~Schieck\cmsAuthorMark{1}, R.~Sch\"{o}fbeck, M.~Spanring, D.~Spitzbart, A.~Taurok, W.~Waltenberger, J.~Wittmann, C.-E.~Wulz\cmsAuthorMark{1}, M.~Zarucki
% ...
\vskip\cmsinstskip
% ...
1:  Also at Vienna University of Technology, Vienna, Austria\\
% ...
\end{sloppypar}
\end{document}